\documentclass{aa}  

\usepackage{graphicx}
\usepackage{tabularx}
\usepackage{amsmath}
\usepackage{appendix}
\usepackage[colorlinks=true,linkcolor=magenta,citecolor=blue,filecolor=cyan,urlcolor=purple,final=true]{hyperref}
\usepackage{txfonts}

\begin{document}

    \title{The effect of data-driving and relaxation model on magnetic flux rope evolution and stability}

   \author{A. Wagner
          \inst{1,2}
          \and
          D. J. Price
          \inst{1}
          \and
          S. Bourgeois
          \inst{3,4}
          \and
          F. Daei
          \inst{1}
          \and
          J. Pomoell
          \inst{1}
          \and
          S. Poedts
          \inst{2,6}
          \and
          A. Kumari
          \inst{5}
          \and
          T. Barata
          \inst{7}
           \and
          R. Erd\'elyi
          \inst{4,8,9}
          \and
          E. K. J. Kilpua
          \inst{1}}

    \institute{Department of Physics, University of Helsinki, P.O. Box 64, FI-00014, Helsinki, Finland\\
    \email{andreas.wagner@helsinki.fi}
    \and
    CmPA/Department of Mathematics, KU Leuven, Celestijnenlaan 200B, 3001 Leuven, Belgium
    \and
    Instituto de Astrof\'isica e Ci\^{e}ncias do Espa\c{c}o, Department of Physics, University of Coimbra, Coimbra, Portugal
    \and
    School of Mathematics and Statistics, University of Sheffield, Solar Physics and Space Plasma Research Centre (SP2RC), Sheffield, United Kingdom
    \and
    NASA Goddard Space Flight Center, Greenbelt, MD 20771, USA
    \and
    Institute of Physics, University of Maria Curie-Sk{\l}odowska, ul.\ Radziszewskiego 10, 20-031 Lublin, Poland 
    \and
    Instituto de Astrof\'isica e Ci\^{e}ncias do Espa\c{c}o, Department of Earth Sciences, University of Coimbra, Coimbra, Portugal
    \and
    Department of Astronomy, E\"otv\"os Lor\'and University, P\'azm\'any P\'eter s\'et\'any 1/A, H-1117 Budapest, Hungary 
    \and
    Gyula Bay Zoltan Solar Observatory (GSO), Hungarian Solar Physics Foundation (HSPF), Pet\H{o}fi t\'er 3., H-5700 Gyula, Hungary}
   \date{Received XXX; accepted XXX}


\abstract
{Understanding the flux rope eruptivity and effects of data driving in modelling solar eruptions is crucial for correctly applying different models and interpreting their results.}
{We investigate these by analysing fully data-driven modelled eruption of the active region (AR) 12473 and AR11176, as well as preforming relaxation runs for AR12473 (found to be eruptive) where the driving is switched off systematically at different time steps. We analyse the behaviour and evolution of fundamental quantities, essential for understanding the eruptivity of magnetic flux ropes (MFRs). }
{The data-driven simulations are carried out with the time-dependent magnetofrictional model (TMFM) for AR12473 and AR11176. For the relaxation runs, we employ the magnetofrictional method (MFM) and a zero-beta magnetohydrodynamic (MHD) model to investigate how significant the differences between the two relaxation procedures are when started from the same initial conditions. In total, 22 simulations are studied. To determine the eruptivity of the MFRs, we calculate and analyse characteristic geometric properties, such as the cross-section, MFR height along with physical stability parameters, such as MFR twist and the decay index. Furthermore, for the eruptive cases, we investigate the effect of sustained driving beyond the point of eruptivity on the MFR properties and evolution.}
{We find that the fully-driven AR12473 MFR is eruptive while the AR11176 MFR is not. For the relaxation runs, we find that the MFM MFRs are eruptive when the driving is stopped around the flare time or later, while the MHD MFRs show eruptive behaviour even if the driving is switched off one and a half days before the flare occurs. We also find that characteristic MFR properties can vary greatly even for the eruptive cases of different relaxation simulations.} 
{The results suggest that data driving can significantly influence the evolution of the eruption, with differences appearing even when the relaxation time is set to later stages of the simulation when the MFRs have already entered an eruptive phase. Moreover, the relaxation model affects the results significantly, as highlighted by the differences between the MFM and MHD MFRs, showing that eruptivity in MHD does not directly translate to eruptivity in MFM, despite the same initial conditions. Finally, if the exact critical values of instability parameters are unknown, tracking the evolution of typical MFR properties can be a powerful tool for determining MFR eruptivity.}

   \authorrunning{A.~Wagner, et al.}
   \titlerunning{The Effect of Data-Driving and Relaxation Model on Flux Rope Evolution and Stability}
   \maketitle

\section{Introduction}
\label{Sect: Intro}

Studying the dynamics of solar magnetic fields is crucial for understanding the initiation and early evolution of solar eruptions, which also contributes to forecasting their space weather effects on Earth \citep[][]{Temmer2021, Kilpua2019, Gopalswamy2022, Buzulukova2022}. The main part of solar eruptions consists of magnetic flux ropes (MFRs), coherent twisted flux tubes with field lines winding about a common axis \citep[][]{Green2018, Chen2017}. Consequently, the behaviour and properties of MFRs have been and continue to be extensively studied in the field \citep[see, e.g.,][and references therein]{Patsourakos2020, Kumari2023, Inoue2023b, Guo2024}{}{}. 

The eruptive behaviour of a coronal MFR is commonly attributed to two instabilities, namely, the kink and torus instability, which in idealised models are triggered when a related stability parameter reaches a critical threshold. For the kink instability, the governing parameter is the field line twist \citep[see, e.g.][]{Torok2004}. For the torus instability, the governing parameter is the decay index $n$, which states how quickly the magnetic field decays with increasing height from the photosphere \citep[see, e.g.][]{Kliem2006}. For both the twist parameter and the decay index, finding the threshold value where the system becomes unstable is highly dependent on the magnetic field configuration of the investigated system \citep[see e.g.,][]{Torok2004, Kliem2006, Schrijver2008}. 

Since the magnetic field of MFRs in the corona cannot be observed directly or remotely with sufficient accuracy, studies of their magnetic configuration and eruptivity often rely on modelling and extrapolations \citep[][]{Green2018}. Magnetohydrodynamic (MHD) simulations have been shown to produce dynamics comparable to observed eruptions 
\citep[see e.g.][]{Fan2007, Aulanier2010, Zhong2021, Torok2024}, they are computationally expensive and often rely on using idealized magnetic field configurations or simplified boundary conditions. One of the more efficient alternative approaches is the magnetofrictional method \citep[MFM;][]{Yang1986} that has also been been widely employed by various authors to model dynamics of solar magnetic fields \citep[][]{Mackay2006, Yeates2014, Price2019, Rice2022}. As the MFM does not use the full equation of motion, it usually dynamically evolve on slower scales.

To capture the dynamics of MFRs from their formation to early evolution for specific observed events, data-driven modelling is paramount \citep[see, e.g.][]{Pagano2019, Pomoell19, Hoeksema2020, Kilpua2021, Inoue2023}. This approach continuously updates the photospheric boundary conditions with new observations, allowing free energy to dynamically build up in the modelling domain. Importantly, there is no need to prescribe an MFR as it can self-consistently form as a response to the evolving conditions at the boundary. For example, the data-driven time-dependent magnetofrictional method (henceforth denoted TMFM) has been successfully used to model MFR eruptions \citep[e.g.,][]{Gibb2014, Yardley2018, Pomoell19, Price2019}. In some studies, magnetic configurations resulting from application of the (T)MFM have also been used as an initial condition in MHD modelling \citep[see e.g.][]{Kliem2013, Pagano2013, Guo2019, Daei2023}.  Although data-driven modelling is evidently a useful tool for investigating MFR eruptions, the effects of driving need to be well understood to correctly interpret the modelling results.

In this work, we utilise a TMFM implementation presented in \cite{Pomoell19} to investigate the MFR formation and eruptivity and how the duration of the driving influences the magnetic field eruptivity. Two active regions are investigated, active region (AR) 12473 and AR11176. The MFR properties are extracted from the simulation data using a recently developed method by \cite{Wagner2024a}, and its corresponding graphical user interface (GUI) called GUI for Identifying and Analysing flux Ropes \citep[GUITAR][]{Wagner2024b}. We consider an MFR as eruptive when it rises through the modelling domain and does not exhibit deceleration. Thus, the eruptivity of the MFRs is scrutinized by investigating the behaviour of its height and size in the simulation domain and contrasting them with the evolution of the aforementioned instability parameters. In the fully driven TMFM runs (i.e., boundary driving is continued for the whole duration of the run) an MFR forms in both cases, but only the one in the former (AR12473) case is found to be eruptive, showcasing also distinct differences compared to AR11176 in how instability parameters evolve throughout the simulation. 

The effect of the driving duration is therefore investigated only for the eruptive AR12473 MFR. We consider both the cases where the driving is stopped before the eruption takes place and when the driving continues while the eruption is already in progress. This is done by systematically varying the time when the boundary driving is stopped and the magnetic field is allowed to evolve without additional stressing (henceforth referred to as relaxation). Furthermore, to investigate whether eruptivity is sensitive to the choice of the physics included in the dynamical model, we also use the magnetic field configuration at the time the driving is stopped as the input to a zero-beta MHD model \citep[][]{Daei2023} in addition to the MFM. A priori, differences between the two are expected, as the approaches differ in particular in the way inertia of the plasma is modeled. However, given that magneto-friction and magnetohydrodynamics are commonly used for coronal modelling, we set out to assess how significant these differences in practise can be.

We note that \cite{Price20} studied the relaxation behaviour of the AR12473 using the TMFM approach for three relaxation times, finding that two of them yielded a rising MFR with torus instability being the most likely cause for the eruption. The present study carries out a more in-depth stability analysis, including different relaxation approaches, and  a finer sampling of relaxation times. 

The paper is structured as follows: In Sect.~\ref{Sect: Methods} we outline the used data, modelling set-up and the FR properties that are calculated, while in Sect.~\ref{Sect: results}, we showcase the properties of the simulated MFRs and their evolution. The results, along with their implications, are discussed in Sect.~\ref{Sect: discussion}. We summarize our findings and answers to the questions in Sect.~\ref{Sect: conclusion}. 

\section{Methods}
\label{Sect: Methods}

\subsection{TMFM model}
This study analyses the magnetic field configurations of NOAA active regions 12473 and 11176. The three-dimensional magnetic field data in a coronal volume surrounding each AR is obtained by employing time-dependent data-driven magnetofrictional (TMFM) simulations. The main characteristic of the model is that the velocity responsible for the dynamics in the corona is proportional to the Lorentz force: 
\begin{equation}
\label{equ: mfm}
\mathbf{v} = \frac{1}{\nu} \frac{\mu_0 \mathbf{J} \times \mathbf{B}}{B^{2}},
\end{equation}
where $\mathbf{B}$ is the magnetic field, $\mu_0$ the vacuum magnetic permeability, $\mathbf{J} = \frac{1}{\mu_0} \nabla \times \mathbf{B}$ the current density, and $\nu$ determines the proportionality between velocity and the magnetic forces, called the magnetofrictional coefficient. The model is initialized with a potential field extrapolation and subsequently evolves the magnetic field in the simulation volume using Faraday's law with a nearly ideal Ohm's law. Changes at the photospheric boundary are being driven by the horizontal electric field inverted from a time series of photospheric magnetograms of the AR \citep[for more details, see][]{Lumme2017, Pomoell19}. Top and lateral boundaries are open, so any structure can leave the domain. The domain extension of the AR12473 simulation is 386 × 254 × 200~Mm, while for AR11176, it is 800 × 374 × 300~Mm. Thus, the domain height of these simulations is 200 and $300\;$Mm for AR12473 and AR11176, respectively. 

\subsection{Relaxation set-up}

As will be detailed in Section 3.2, an MFR forms in the TMFM simulations of the ARs, but only for the case of AR12473 does the model exhibit behaviour indicative of an eruptive structure, i.e. as stated in the Introduction the MFR rises without deceleration, and finally, starts exiting the modelling domain. Therefore, the simulation for AR12473 allows to investigate the point in time when the magnetic field configuration exhibits eruptive behaviour. 

This analysis is performed by running a set of additional model runs, denoted as relaxation runs, where the photospheric driving is switched off at a specific time $t_0$, i.e., the horizontal electric field responsible for evolving the photosphere is set to zero beyond $t_0$, $\mathbf{E}_h(x, y, z=0, t >= t_0) = 0$. We systematically vary the relaxation time $t_0$ and investigate how the MFR subsequently evolves, given different initial magnetic field configurations, representing different stages of its fully-driven evolution. 

\begin{figure}
     \centering     \includegraphics[width=\linewidth]{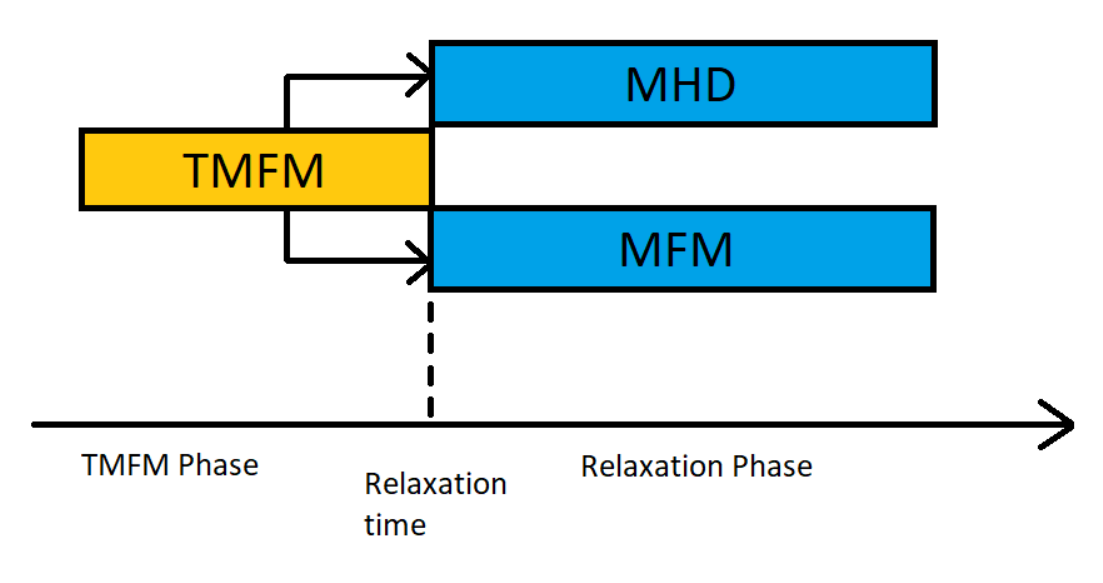}
     \caption{Schematic for our simulation set-up, showing the relaxation procedure. For the fully-driven cases, the yellow bar runs until the end of the blue bars.}
     \label{fig: Schematic}
\end{figure}

Furthermore, we evolve the system after time $t_0$ using two distinct modelling approaches (illustrated in Fig~\ref{fig: Schematic}): in approach 1) we continue the computation employing the magnetofrictional prescription, whereas in approach 2)  we employ a zero-beta MHD model from \cite[][]{Daei2023}. The zero-beta approach relies on the assumption that the coronal plasma is magnetically dominated, such that thermal pressure and gravity are negligible. The initial flow speed is set to zero, while the initial density is distributed such that the Alfv\'en speed is constant in the entire domain. With this prescription, the initial dynamics in the zero-beta MHD model are proportional to the TMFM speed and thus the Lorentz force. More details on the numerical aspects and transfer from the initial TMFM conditions to the MHD model are presented in \cite{Daei2023}.

Utilizing both the MFM and zero-beta MHD relaxation approach enables us to investigate whether the eruptivity of a given configuration is sensitive to the chosen physical model and if so, how significant the differences are. All models use the exact same spatial resolution, as detailed in \citep{Daei2023}. 

We run 9 MFM and 11 MHD relaxations in addition to the two fully driven TMFM simulations, making 22 simulations to analyse. The fully driven AR12473 TMFM simulation runs from 22 Dec 2015, 23:36 to 2 Jan 2016, 12:36 \citep[][]{Price20}. The relaxation times span from 27 Dec 2015, 12:36 to 29 Dec 2015, 12:36 (with a cadence of six hours between the different relaxation times) for MFM and from 27 Dec 2015, 00:36 to 29 Dec 2015, 12:36 for the MHD relaxation (using the same cadence of relaxation times). The AR11176 TMFM run spans from 25 Mar 2011, 4:00 to 1 Apr 2011, 18:00. 

To simplify the notation, we shorten the notation of the starting time of a given relaxation run to include only the day and hour, e.g., 27 Dec 2015, 12:00 becomes "27-12". The relaxation set-ups are summarized in Table~\ref{table:configs}. All (T)MFM simulations are analysed on a cadence of six hours.

\begin{table*}
\caption{Overview of the relaxation runs of AR12473.}
\begin{tabularx}{\textwidth}{|| >{\centering\arraybackslash}X 
| >{\centering\arraybackslash}X 
| >{\centering\arraybackslash}X 
| >{\centering\arraybackslash}X ||} \hline \label{table:configs}
Acronym & Relaxation Time & MFM & 
MHD \\
\hline\hline
27-00 & 27 Dec 2015, 00:36 &  & X \\
\hline
27-06 & 27 Dec 2015, 06:36 &  & X \\
\hline
27-12 & 27 Dec 2015, 12:36 & X & X \\
\hline
27-18 & 27 Dec 2015, 18:36 & X & X \\
\hline
28-00 & 28 Dec 2015, 00:36 & X & X \\
\hline
28-06 & 28 Dec 2015, 06:36 & X & X \\
\hline
28-12 & 28 Dec 2015, 12:36 & X & X \\
\hline
28-18 & 28 Dec 2015, 18:36 & X & X \\
\hline
29-00 & 29 Dec 2015, 00:36 & X & X \\
\hline
29-06 & 29 Dec 2015, 06:36 & X & X \\
\hline
29-12 & 29 Dec 2015, 12:36 & X & X \\
\hline
\end{tabularx}
\end{table*}

Note that we cannot directly compare MHD relaxation runs with MFM relaxation runs or the fully driven TMFM runs because the time scales of dynamics in these two modelling approaches differ (see also \citep[][]{Wagner2023}). The MFM simulations are run until the same final time as in the fully driven simulation (i.e., until 2 Jan 2016, 12:36), while for the MHD simulations, we analyse the results over a time span during which the magnetic field undergoes a qualitatively similar evolution (corresponding to the first 30 outputs after the start of the relaxation).

\subsection{Flux rope extraction scheme}
\label{Sect: Extraction}
A method to extract the relevant portion of the three-dimensional magnetic field is needed in order to analyse the properties of the MFRs in the model output. In this work, we use the MFR extraction and tracking tool GUITAR \citep[Graphical User Interface for Tracking and Identifying flux Ropes][]{Wagner2024b}, which follows the scheme presented in \cite{Wagner2024a}. We briefly summarize the algorithm: First, the twist number $T_w$ is calculated in the modelling domain via a code adapted from \cite{Liu16} and evaluated in a 2D slice (we refer to the resulting maps as twist maps). The twist number is defined as follows: 
\begin{equation}
\label{equ: twist}
  T_w = \int_{L} \frac{\mu_0 J_{\parallel}}{4 \pi B} \, dl,
\end{equation}
with $J_{\parallel}$ the component of the current density parallel to the magnetic field and $L$ the field line of interest. It quantifies the amount that two infinitesimally close field lines turn about each other \citep{Berger2006} and is a commonly used approximation for the winding number \citep{Liu16}. The plane is chosen (visually) for the 2D slice such that it cuts through the MFR (see \cite{Wagner2024a} for details on the geometry of the cutting planes for the AR12473 and AR11176 runs). Alternatively, one could also calculate the winding number $T_g$, which measures the winding of magnetic field lines about a chosen axis. The calculation of $T_g$ is, however, non-trivial \cite[e.g.][]{Price2022} and as large amounts of twist maps are needed in this study, we opted to use $T_w$. 

Due to the varying complexity in the evolution of $T_w$ during the course of the simulation we make use of Mathematical Morphology (MM) methods in order to extract the MFR throughout the time series. These image processing and feature recognition tools compare image elements to structuring elements (SEs), which are predefined geometrical shapes. The erosion and the dilation algorithm are two fundamental procedures of the MM toolset. The erosion of a binary image $X$ is defined as $\epsilon_{A}(X) = X \ominus A$, which is the collection of all centre points $a$ of the SE $A$, such that $A$ is fully included in $X$. The dilation, on the other hand, is defined as $\delta_{A}(X) = X \oplus A$, which is the union of all the centre points $a$ of the SE $A$, such that $A$ and $X$ have non-zero overlap. Finally, these algorithms can be combined in various ways, for example, to form the opening algorithm: $\gamma_{A}(X) = \delta_{A}(\epsilon_{A}(X))$. 

We use them, for example, to sharpen the features of the twist maps to stabilize the thresholding procedure (morphological gradient) and separate connected features in the maps (opening algorithm). After the MM processing, a threshold is applied to identify a set of seed points of the magnetic field lines. We note as some MFRs were inherently more twisted than others, we applied the following threshold ranges: $T_w^{FR} \in [0.6; 0.8]$ for the relaxation runs, $T_w^{FR} = 0.8$ and $T_w^{FR} = 0.4$ for the fully driven TMFM runs for AR12473 and AR11176, respectively. The resulting shapes (i.e., the MFR cross-sections in the chosen plane) are associated based on their overlap between consecutive frames. This procedure lets us track the evolution of a particular MFR throughout the simulation. Subsequently, we sample source points from the extracted MFR regions and visualize their magnetic field lines. Further details on the method are provided in \cite{Wagner2024b}. 

\subsection{Stability properties}

To gain insight into potential reasons for the eruptivity of the MFRs, we analyse the twist number and decay index, the quantities of importance for triggering the kink instability (twist number) and the torus instability (decay index). The twist number $T_w$ is directly obtained as part of the MFR extraction scheme (see Equation \ref{equ: twist} in Sect.~\ref{Sect: Extraction}). 
 
The distribution of $T_w$ can vary inside a given MFR, thus, we compute the average of $T_w$ over all field lines contained in the MFR \citep[similar to, e.g.,][]{Zhong2021, Duan2022}. As discussed in the Introduction, the critical value for the twist number ($T_w^c$), when the MFR is expected to become kink-unstable, depends on the MFR geometry and magnetic configuration \citep[see, e.g.,][]{Torok2004}. Therefore, we focus on evaluating the evolution of this quantity during the course of the simulation. However, we provide a reference value in the figures of $T_w = 1.25$, which is the critical number of field line turns for a force-free cylindrical MFR with uniform twist, derived by \cite[][]{Hood1981}. 

As discussed in Sect.~\ref{Sect: Extraction}, a more exact quantification of the MFR twist is the winding number $T_g$. However, due to the large number of simulations, $T_w$ was chosen over $T_g$ for computational efficiency. Furthermore, $T_w$ and $T_g$ are known to follow similar trends, aside from possible systematic differences in their magnitudes \citep[see e.g.][who found a systematic offset of 0.4 turns between the two]{Duan2022}. Since our focus here is on the evolutionary trends of the instability parameters, this difference does not affect our study. 

To investigate the behaviour of the magnetic field surrounding the MFR, we calculate the decay index $n$ as follows: 
\begin{equation}
\label{equ: decay index}
n = -R\frac{d\ ln(B_{ex})}{dR},
\end{equation}
where $R$ is the height above the photosphere and $B_{ex}$ is the external (i.e., surrounding the MFR) magnetic field. The external magnetic field $B_{ex}$ is estimated by calculating the overlying magnetic field via a potential field extrapolation using the photospheric magnetic field data \citep[see, e.g.,][]{Zuccarello2014, Guo2010}. We evaluate this parameter in the same plane as the $T_w$-maps in our extraction scheme, using the horizontal components of the magnetic field ($B_x^{\mathrm{pot}}$ and $B_y^{\mathrm{pot}}$) from the potential field. To estimate the decay index affecting the MFR, we calculate $n$ near the MFR centre. In more detail, we take the value of $n$ at the centre of the MFR cross-section in the plane of our twist maps. We compared this value against $n$, an average over a rectangle centred around the MFR midpoint, with the width and height chosen to be one-fifth of the corresponding dimensions of the cross-section of the MFR. Since the values obtained using these two approaches are identical to within 2 $\%$, we conclude that there are negligible variations of $n$ close to the MFR centre (cf. \ref{Appendix}). Thus, we use the value of $n$ at the MFR centre in our analysis. Similarly, as for the twist parameter, we indicate a reference using a value of $n = 1.5$ - the limit for a circular, thin current channel \citep[][]{Demoulin2010}, in agreement with previous studies \citep[e.g.,][]{Zuccarello2016, Sarkar2019}. 

In addition to the magnetic field parameters, we calculate the evolution of the geometric parameters of the MFR, specifically, the cross-section and height in the twist map planes. This allows us to track how MFRs grow and move through the simulation domain. 

\section{Results}
\label{Sect: results}

\subsection{General MFR appearance}

\begin{figure*}
     \centering     
     \includegraphics[width=0.9\linewidth]{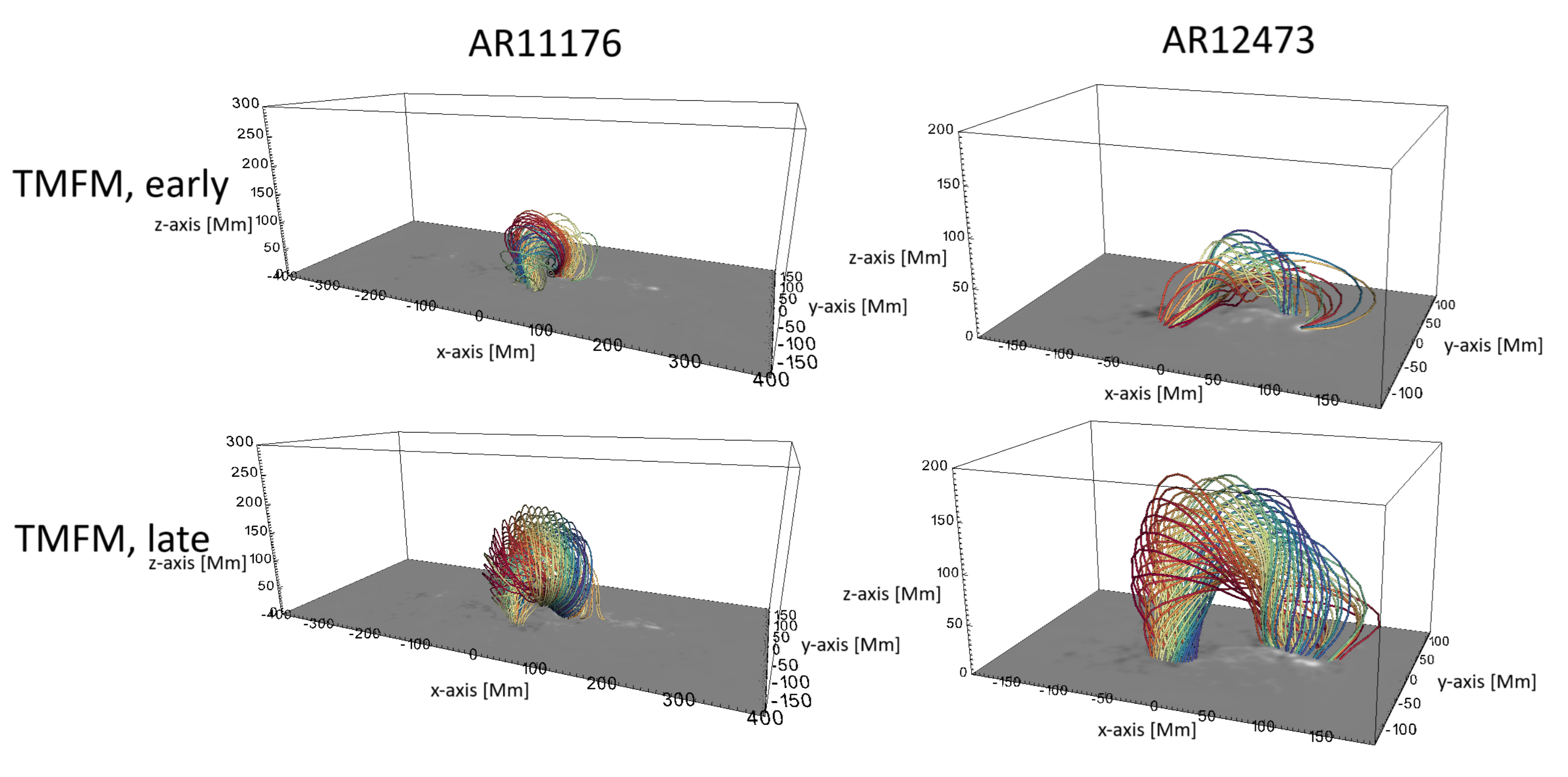}
     \caption{Characteristic snapshots of TMFM run of AR12473 (top) and AR11176 (bottom). Field line colouring is based on seed point location, chosen for appropriate contrast. The $B_z$ component of the magnetic field is plotted at the bottom boundary, with the conventional colour grey-scale scheme (positive polarity in white, negative polarity in black.  The respective maximum and minimum colour values are at 0.3T and -0.3T).}
     \label{fig: TMFM FR Evol}
\end{figure*}

An overview of the magnetic field structure of the extracted MFRs of the fully driven TMFM runs is shown in Fig.~\ref{fig: TMFM FR Evol}. The AR12473 MFR forms a very coherent structure, but as already found in \cite{Wagner2024b}, it consists of at least two entangled MFRs. This can be best seen from the early-stage snapshot in Fig.~\ref{fig: TMFM FR Evol} where a set of field lines that connects to the same region of negative polarity $B_z$ is rooted in a different region of positive $B_z$ than the main bulk of field lines. As the MFR evolves, these two structures progressively merge, which is evident from the migration of the inner positive footpoints towards the outer ones. Furthermore, the MFR appears to be erupting, as it moves toward the top boundary and eventually starts to exit the simulation domain. 

The AR11176 MFR also appears coherent and consists of two main structures: a core bundle of twisted field lines enveloped by less twisted field lines around the apex region. As the simulation progresses, the MFR visually appears to rise, which is primarily due to expansion rather than an increase in height. Especially, the envelope grows larger and gets progressively twisted, while the twist in the core progressively drops, consistent with our previous findings \citep[cf. Fig~5 in][]{Wagner2024a}. 

\begin{figure*}
     \centering     \includegraphics[width=\linewidth]{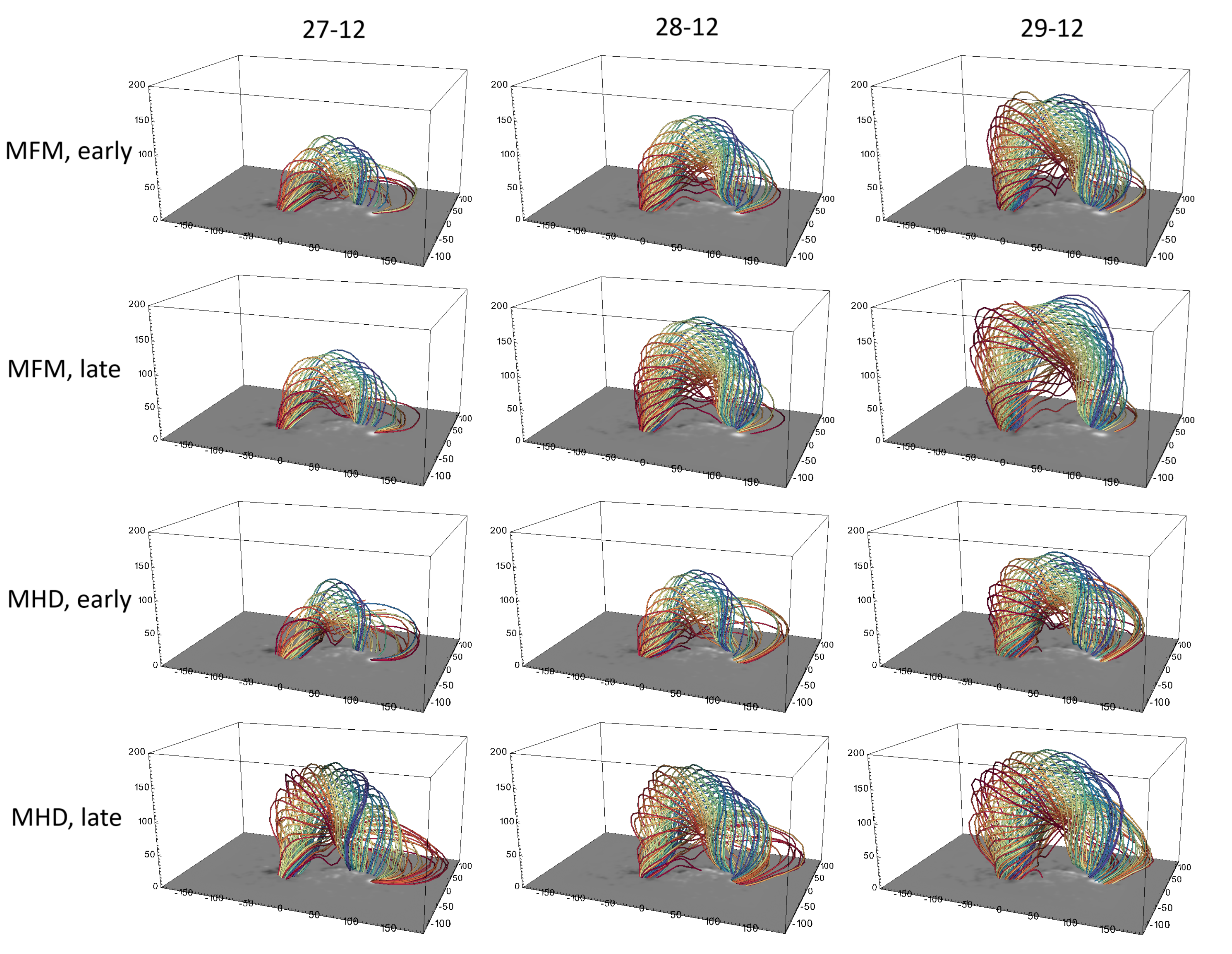}
     \caption{Characteristic snapshots of MFM and MHD relaxation runs of AR12473. The right column shows runs initiated with the TMFM state on 27 Dec 2015 at 12 UT, the middle column is initiated with the TMFM state on 28 Dec 2015 at 12 UT and the left column is initiated with the TMFM state on 29 Dec at 12 UT. The first two rows display the early and late stages of the MFM relaxation MFR, while the last two rows show the early and late stages of the MHD relaxation MFR. Axes, viewing angle and $B_z$ colouring are equivalent to Fig.~\ref{fig: TMFM FR Evol}}
     \label{fig: FR Evol}
\end{figure*}

A set of representative snapshots from the relaxation simulations is presented in Fig.~\ref{fig: FR Evol}. The MFM MFRs appear slightly more coherent than the MHD MFRs and notably less twisted, especially in the later stages. The MFM MFRs rise higher the later the chosen relaxation time is. The MFR in this early relaxation run barely changes height while the two other examples with the relaxation taking place at later times rise (see top 2 rows of Fig.~\ref{fig: FR Evol}). Furthermore, the MFRs in the runs using later relaxation times grow thicker compared to the MFR in the run using the earliest relaxation time.
Compared to the MFM relaxation runs, the MFRs extracted from the MHD relaxation runs appear significantly more twisted and less coherent, though with similar sub-structures: There is yet again a set of field lines present that connects to a different region of positive polarity magnetic field at the photospheric level, which is present throughout the whole MFR evolution. The most striking difference, however, is that the height reached by the MFRs in the MHD relaxation simulations does not depend as clearly on the chosen relaxation time.  

\subsection{Fully driven MFM simulation runs}
\label{Sect: TMFM}

\begin{figure}
     \centering     \includegraphics[width=\linewidth]{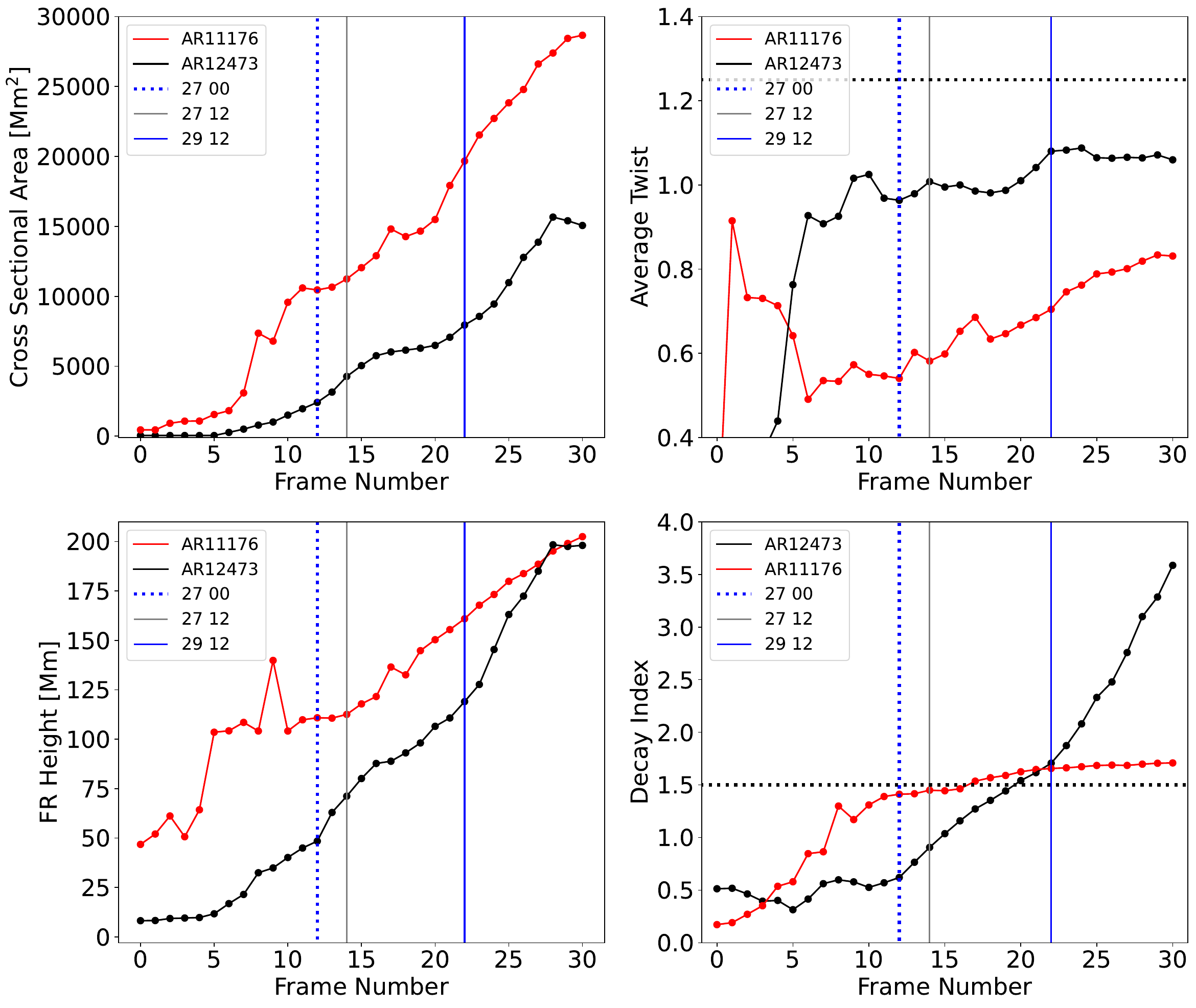}
     \caption{MFR Properties for the fully-driven TMFM MFRs of AR12473 (black) and AR11176 (red). The AR12473 relaxation windows are indicated with vertical lines. The MFM relaxation runs start at 27 12 (grey line), while the MHD runs start at 27 00 (dotted blue line). The latest relaxation time for both cases is 29 12, marked by the blue vertical line. Dotted black horizontal lines highlight reference values of $T_w = 1.25$ and $n = 1.5$ in the respective panels.}
     \label{fig: FR TMFM}
\end{figure}

The evolutionary trends of the cross-sectional area, apex height (calculated here as an average of the field lines within 5 Mm of the "true apex" to minimize the effect of outliers), average twist and decay index of the magnetic flux rope in the fully driven simulation are shown in Figure~\ref{fig: FR TMFM}. The MFR cross-sectional area for AR12473 (black curve) in the cutting plane (see Section~\ref{Sect: Methods}) grows monotonically until around 28th Dec 2015 at 06:00 UT (frame 17), after which it first tentatively plateaus and then rises again until the MFR approaches the top of the modelling domain. At this point, the growth of the cross-section abruptly stops and declines as the MFR starts leaving the simulation domain. A qualitatively similar trend is observed for AR11176 (red curve), with the MFR more or less continuously growing until the end of the simulation. This figure demonstrates that the AR11176 MFR is significantly thicker than the AR12473 MFR throughout the simulation. 

The average $T_w$ of the AR12473 MFR first rises sharply during the formation phase and stabilizes afterwards with only a moderate increase towards the end of the simulation. However, the reference line set at $T_w = 1.25$ is not reached, remaining at values below 1.1 turns. There are some notable fluctuations such as the extended dip that occurs within the time window of the chosen relaxation times. 
The AR11176 MFR features a strong early peak of average $T_w$ but falls to low values of almost 0.5. This peaking and subsequent drop of average $T_w$ likely results from our detection method, which only captures the twisted MFR core in the early stages (cf., \cite{Wagner2024a} and the twist map evolution of AR11176 in the supplementary material), while it also captures the weakly twisted surrounding fields at the later phases, which get gradually more twisted. 
At this point (frame 6) $T_w$ then rises again but stays below 0.8, while also the rate at which $T_w$ increases falls off. \\

The height of the AR12473 MFR increases monotonically, similar to the cross-sectional area. The saddle point-like feature is also visible at the same time step but is less pronounced than in the cross-section area evolution. Towards the end of the simulation, the MFR apex rise stops at about 200~Mm when the top of the modelling domain is reached. The AR11176 MFR height evolution, in turn, features a sudden jump in the early stages (similar to the above, related to the MFR extraction method "suddenly" including surrounding twisted fields). Except for a few outliers, the MFR continuously rises, reaching 200~Mm at the end of the simulation (note that the domain height for this particular simulation is 300~Mm). It is however evident from Fig~2 (as well as Fig~5 in \cite{Wagner2024a}) that the increasing height is attributed to the MFR growth, rather than the whole structure lifting off. \\

Finally, Fig.~\ref{fig: FR TMFM} shows the decay index at the MFR centre over time for both fully driven simulations. The AR12473 MFR forms at around $n = 0.5$. First, the decay index remains roughly stable around this value, but as the MFR grows and rises, $n$ increases linearly. At the later stages, close to the latest chosen relaxation times, the decay index increases rapidly, still exhibiting linear behaviour but with a considerably steeper slope than earlier. The AR11176 MFR shows a quicker initial rise in $n$, which aligns with the evolution of the other studied parameters. However, when the indicated reference line $n = 1.5$ is approached, the evolution of $n$ stabilizes with only a weak increase until the end of the simulation. This starkly contrasts with the AR12473 MFR, which experiences a substantial increase in $n$. 

\subsection{Magnetofrictional relaxation runs}
\label{Sect: MFM}
\begin{figure}
     \centering     
     \includegraphics[width=\linewidth]{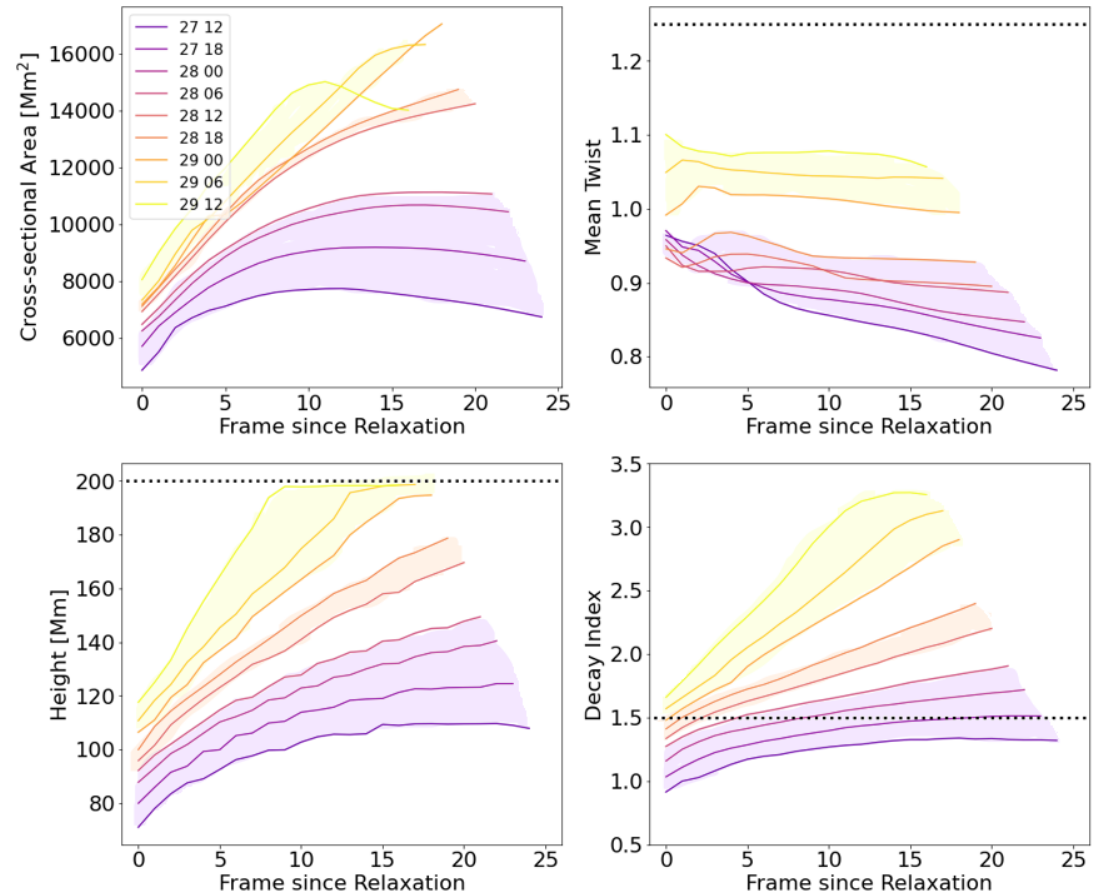}
     \caption{MFR properties for the MFM relaxation runs of AR12473. The colours of the curves indicate the time at which the relaxation is started, where blue/purple indicates that the relaxation start is chosen at earlier times from the fully driven case, and correspondingly yellowish colors indicate a later starting time. MFRs that evolve similarly (quantitatively and/or qualitatively) have a shaded background. The starting point of each curve is synchronized to the relaxation time.}
     \label{fig: FR Props MFM}
\end{figure}

Figure~\ref{fig: FR Props MFM} shows the same quantities as a function of simulation frames for the AR12473 MFM relaxation runs as were given in Figure~\ref{fig: FR TMFM} for the fully driven runs. Here, the data is plotted as a function of time from the start of the relaxation (which varies from case to case), thus showing how each initial condition evolves from the time of the start of the relaxation. Note that the simulation has been run in all cases until the same end time as the fully driven run. 
As a result, each MFM relaxation run has a different duration and total frame count, with the temporal cadence of the output chosen to be identical for all runs.

The figure clearly shows that the MFR parameters evolve in time rather smoothly without abrupt changes. Firstly, the top left panel shows that the MFR cross-sectional area is larger the later the relaxation time is chosen, with only a few exceptions. Naturally, those MFRs that reach the top of the simulation domain before the simulation ends show a significant decrease in their cross-section from that point onwards. \\

In contrast to the fully driven case, the average twist of the MFRs in the MFM relaxation (top right panel) either decreases for the whole modelling duration (earlier relaxation times) or first increases and then decreases (later relaxation times). Mostly, the variations during the runs are small, especially for the middle and later relaxation times, for which the twist value stays approximately constant in time. We point out that the latest three relaxation times produce MFRs that showcase a notably higher average twist and are the only ones reaching $T_w > 1$. However, none of the MFM relaxation MFRs reach the reference line set at $T_w = 1.25$, with the highest values being $T_w \approx 1.1$ for the latest (29-12) MFM relaxation run. \\

The height evolution, shown in the bottom left panel, indicates that all mid and late-stage relaxation times produce a rising MFR (curves with orange- and yellow-shaded background in Fig.~\ref{fig: FR Props MFM}). The earlier relaxation times show a stagnation towards the end, indicating that the MFR evolution might stabilize with the MFR settling to a given height (purple shaded background). The height for the earliest relaxation time (27-12) shows even a slight decreasing trend in the last frames. Interestingly, the time at which the height of the early-stage relaxation time MFRs start to stagnate roughly coincides with the decrease in the cross-sectional area. 

Lastly, the decay index evolution, displayed in the bottom right panel, shows an approximately linear behaviour for the late-stage relaxation runs, while the mid and early stage ones show a decelerating increase that eventually ends in a linear phase as well. It is again notable that the curves for the three latest relaxation times (29-00, 29-06, 29-12, i.e., yellow shaded areas in Fig.~\ref{fig: FR Props MFM}) have considerably steeper slopes in their linear phase than the other cases. Note that the linear evolution is halted by the MFRs encountering the top of the simulation domain, after which their central point changes only marginally. 
However, only the earliest MFM relaxation MFR (27-12) stays clearly below the reference line of $n = 1.5$, while the 27-18 run MFR barely reaches it in the very late stages of the simulation. For all other cases, the decay index curves are clearly above the reference line, with those corresponding to the two latest relaxation times placed above it from the beginning. 

\subsection{MHD relaxation runs}
\label{Sect: MHD}

\begin{figure}
     \centering
     \includegraphics[width=\linewidth]{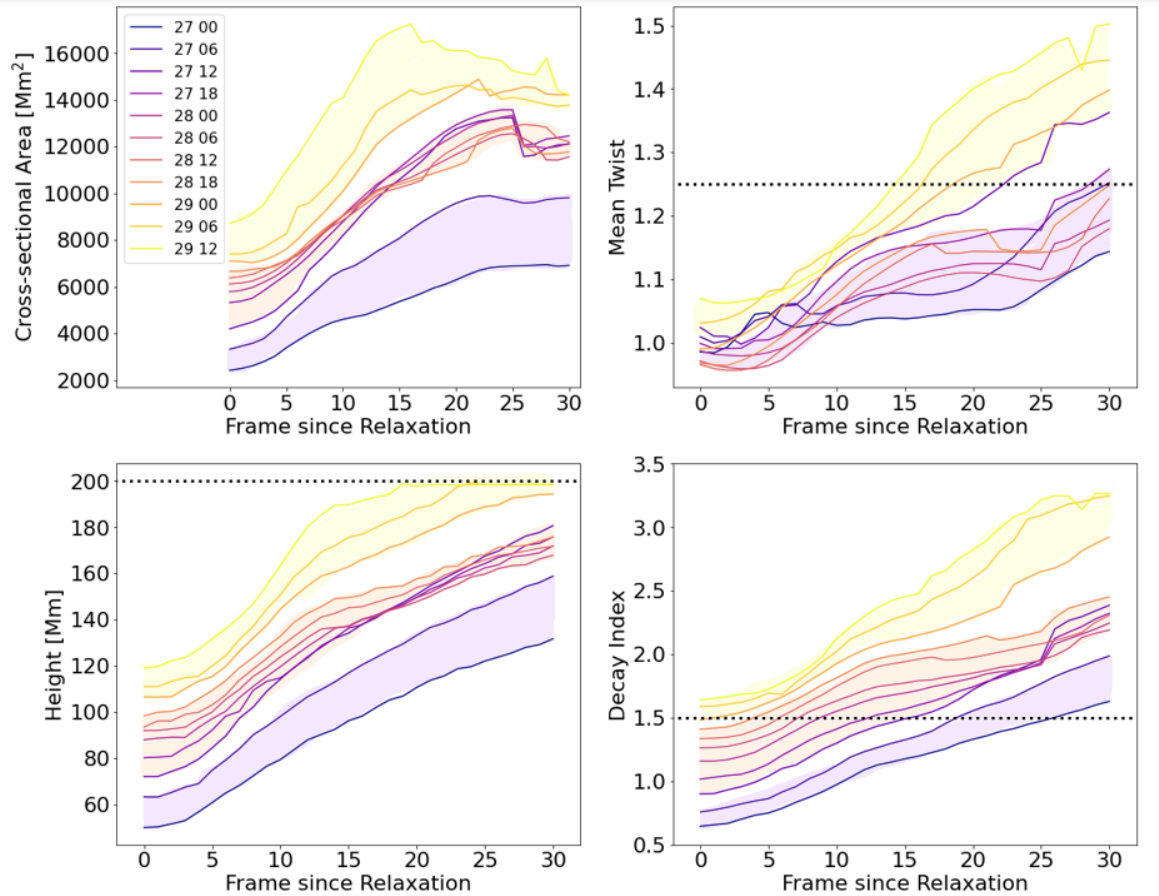}
     \caption{Same as Fig.~\ref{fig: FR Props MFM}, but for the MHD relaxation run MFRs of AR12473.}
     \label{fig: FR Props  MHD}
\end{figure}

The properties of the MFRs in the MHD relaxations, displayed in Figure~\ref{fig: FR Props  MHD}, show more erratic evolution compared to the MFM simulations. For example, for some cases, the evolution of the cross-section appears to exhibit a jump that is a result of the MFR encountering the top of the domain (cf., height evolution in the panel below) or from the extraction procedure. Here, to retain the coherence of the MFR, it was necessary to cut off some previously attached MFR features that stopped following the movement/evolution of the main large-scale structure from the twist maps by using a series of MM algorithms (opening). 

The cross-sectional MFR area curves in the selected extraction plane (see Section~\ref{Sect: Methods}) show a (quantitatively) large separation between the latest and earliest relaxation time MFRs throughout the simulation. For the runs with relaxation start times falling between 27-12 and 28-18, in contrast to MFM relaxation results, the cross-section area curves are clustered together following a very similar evolution  (cf. orange shaded areas). However, most cross-sectional area curves for MHD relaxation runs exhibit a similar trend regardless: slow increase or stagnation during the first frames, followed by an extended period of rapid and approximately linear increase that then levels off. The slopes in the linearly increasing phases get steeper the later the relaxation time is chosen. 

Except for a brief stagnation at the start of each simulation, the average MFR twist curves in the MHD relaxation simulations show a predominantly increasing trend for all relaxation times. This is in strong contrast with the mainly decreasing trends found for the MFM relaxation runs. Similarly to the MHD cross-sectional area, the average twist is not entirely organized with the relaxation time choice, particularly in the mid-range. Furthermore, for the three latest as well as the 27-12 relaxation time runs, the mean twist curves reach above the reference line of $T_w = 1.25$. A few other curves also reach or slightly pass the reference curve in the later stages of their respective simulation. 

The height evolution shows similar trends to the cross-sectional area (until the possible encounter with the top boundary). 
First, there is a short phase where the MFR height stagnates. After this initial phase, the MFRs rise rapidly, followed by a phase where the increase in height becomes linear with a relatively steep slope. Eventually, they transition into another linear phase with a shallower slope. This last phase tends to be more distinct for later relaxation times. It is also notable that the duration of the rapid rising phase is shorter for the earliest relaxation times (27-00 and 27-06). The slopes of the linear phases also differ. For example, the mid-relaxation times (between 28-00 and 28-18) have particularly shallow slopes (lower acceleration) in the late stages than earlier ones, and  27-12 and 27-18 MHD MFRs eventually overtake them. \\

The decay index evolution appears similar to the early stages' height and cross-sectional evolution curves. They also exhibit the same behaviour as observed for the other parameters in both MFM and MHD runs, being that there is a notable gap between most of the runs and the properties of the MFRs in the latest three relaxation runs. This is highlighted with the yellow background colouring in Fig.~\ref{fig: FR Props  MHD}. Furthermore, the mid-stage relaxation times cross with the early-stage relaxation time curves in the later stages of the simulation - except for the two additional very early-stage MHD runs. In contrast to the MFM decay index evolution curves, all MHD-run MFRs eventually surpass the reference line of $n = 1.5$. Additionally, not even the earliest relaxation times saturate at a constant level, as is the case for the MFM runs.

\begin{figure}
     \centering
     \includegraphics[width=\linewidth]{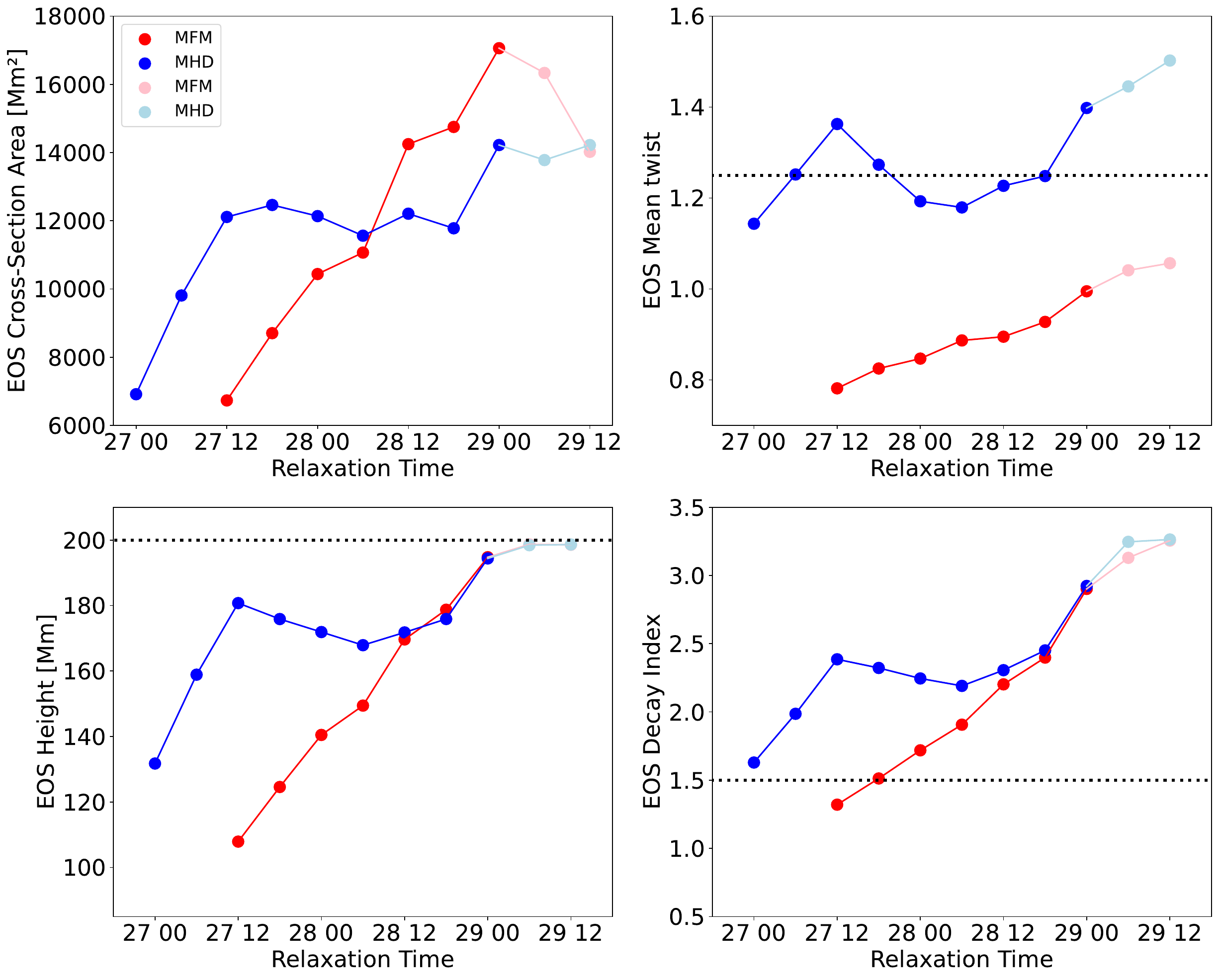}
     \caption{MFM (red) and MHD (blue) relaxation MFR properties at the end of the simulation (EOS). The ending time corresponds to the ending time of the fully-driven simulation for MFM, while for MHD, it corresponds to the 30th frame after relaxation. Points for MFRs, which interact or exit the domain at the end of the simulation have been made transparent.}
     \label{fig: FR Final Props}
\end{figure}

\subsection{Comparison of end-of-simulation properties}
\label{Sect: EOS Props}
To understand how the investigated MFR properties evolve, the final values at the end of the simulation (EOS) as a function of the chosen relaxation start time is shown in Fig.~\ref{fig: FR Final Props}. For MFM runs, all parameters show an approximately linearly increasing trend, i.e. reaching higher values the later the relaxation time is chosen. Exceptions are simulations in which the MFR reaches the top of the domain and starts to exit, directly influencing the MFR cross-section and caps off the MFR height at about 200~Mm (cf. transparent data points). As already suggested by Fig.~\ref{fig: FR Props  MHD}, the MHD relaxation EOS properties show a more complex behaviour than the runs employing MFM relaxation: They increase only until the relaxation time 27-12, where the curves have a local maximum (for the cross-sectional area this occurs at 27-18). The curves then dip, achieving their local minimum at 28-06. Afterwards, the properties increase again, although saturation again sets in for those MFRs that reach the top of the domain.

Interestingly, the evolution of MHD relaxation EOS quantities resembles the evolution of the average twist profile of the TMFM simulation in Fig.~\ref{fig: FR TMFM} in the relaxation time window. It features the local maximum at 27-12 followed by an extended dip, after which $T_w$ increases again. However, the local minimum here is at 28-12 (frame 18 in Fig~\ref{fig: TMFM FR Evol}). This point also corresponds to the saddle point in the height and cross-sectional area evolution. 

\section{Discussion}
\label{Sect: discussion}

In this work, we employed an extraction method to analyse the evolution of the magnetic flux ropes (MFRs) in the modelling results, developed in \cite{Wagner2024a}. We note that the results further demonstrate that the scheme is suited for time-efficient analysis of large batches of simulation data. Manual post-processing was used only in cases where it was absolutely necessary and was done in a systematic way by applying  mathematical morphology (MM) algorithms, such as the morphological gradient and the opening algorithm. 

\subsection{Flux rope properties in the TMFM run}
The MFRs extracted from the fully driven TMFM simulations of AR12473 and AR11176 (properties depicted in Figs.~\ref{fig: FR TMFM}) show a continuous increase in the geometric properties (cross-sectional area, height) as well as the magnetic field instability-related properties except for an early spike in $T_w$ for the AR11176 run. This initial spike is due to the small size of the MFR at the beginning and the way it was formed. As discussed in \cite{Wagner2024a}, the MFR in this event forms at a height of approximately 50~Mm from initially untwisted field. After formation, field lines surrounding the core get gradually more twisted. Consequently, the extraction algorithm first captures solely the highly twisted core, resulting in a $T_w$ spike, after which when surrounding (initially significantly less twisted) fields are recognized as part of the MFR, the average $T_w$ drastically drops. Later, $T_w$ gradually increases when the twist in the surrounding field increases. The nearly monotonous slow rise of the MFR from about 100~Mm onwards, after its complex initial phase, suggests that the MFR has fully formed at this stage (frame 10). The AR12473 MFR, in turn, forms quickly and at considerably lower heights. Other crucial differences between the MFRs in AR11176 and AR112473 are the rate at which the values of the investigated properties increase and the final values attained. In particular, the height and decay index for AR11176 MFR increase notably slower. It is important to note that the increase in height of the AR11176 MFR is primarily due to its strong expansion, as evident from the cross-sectional area evolution, rather than a lift-off from the surface. The AR11176 MFR is also notably less twisted than the AR12473 TMFM MFR. Combined, these findings imply that AR11176 MFR was non-eruptive, while there is strong evidence that the MFR in AR12473 erupts and that torus instability is the most probable cause for the eruptivity.

Most notably, in the later stages of the simulation, the MFR height and cross-sectional area increase fast, along with a rapidly rising decay index well past the reference value. The slow increase of $T_w$ and its value staying clearly below the reference level suggest that kink instability playing an essential role in the eruption is unlikely.

\subsection{Flux rope properties in the MFM relaxation runs}
The MFM relaxation MFR properties of AR12473, displayed in Fig.~\ref{fig: FR Props MFM}, show a remarkably smooth evolution. The curves in the figure are also ordered according to the relaxation times, which can also be seen in the properties at the end of the simulation (EOS), in Fig.~\ref{fig: FR Final Props}. 

Some distinct differences in the behaviour of the curves are, however, clear. The cross-sectional area for the three latest MFM relaxation MFRs displays a strong linear increase throughout the simulation until the domain boundary is reached. The MFRs in the earlier runs begin with a comparable growth rate but their growth decelerates as the simulation progresses, and for the earliest relaxation time, turning even into a decrease in the cross-sectional at the later stages. These systematic differences are marked with different background colours in Fig.~\ref{fig: FR Props MFM}.

The MFM MFR twist, in turn, stays approximately at the same level or decreases. This decreasing trend might be tied to the nature of the model, which evolves towards a minimum-energy force-free state according to \cite{Yang1986}. Thus, eruptivity is only expected when the MFRs have been driven to a point where the most energy-efficient path of evolution is an eruption. As the average $T_w$ does not exhibit explosive behaviour in the fully driven TMFM case, it is an expected result that such behaviour does not occur in the MFM relaxation runs either.   
 
All MFM relaxation MFRs start rising approximately linearly at the start of the simulation. The linear increase in height is sustained, however, only for the relaxation simulations that are started at later times (28-12 and later). These are the MFM MFRs that we consider to be eruptive. The MFRs in the relaxation simulations initiated earlier in the evolution of the AR show deceleration already in the early to mid stages of the simulation. This quasi-stabilization coincides with the time that their cross-sectional areas stabilize and start to decrease. The runs initialized from late and mid-stage relaxation times have a decay index (28-12 onward) values reaching at least about $n = 2$. Combined with the observed trends in the evolution of cross-section and height, this is a strong indicator that this set of MFRs erupt due to the torus instability. In contrast, the MFRs in the relaxation runs started at earlier times in the AR evolution were likely not in a state that favours eruption in the MFM approach. The trends and attained values in the computed average $T_w$ profiles indicate that as in the TMFM case, the kink instability does not play a crucial role in destabilizing these MFRs.
We note that while some studies suggest that a combination of kink and torus instability may be necessary to destabilize specific magnetic field configurations \citep[][]{Myers2015, Zhong2021}, this is not a general requirement \citep[see, e.g.,][]{Fan2007, Aulanier2010, Kliem2013}. The above results are consistent with those presented in \cite{Price20}.

Furthermore, a "phase transition" is visible in Fig.~\ref{fig: FR Props MFM}, i.e., there is a significant gap caused by an abrupt change in the behaviour between two subsequent evolutionary curves for relaxation times 28-18 and 29-00 (cf. coloured background). This is particularly apparent for cross-sectional area, twist and decay index. However, both the 28-18 MFM MFR and the 29-00 MFM MFR exhibit eruptive behaviour, demonstrating that continued driving past the eruptive point may still significantly influences the MFR evolution.  

\subsection{Flux rope properties in the MHD relaxation runs}
The MFR properties derived from the MHD runs (height, twist, cross-section and decay index), displayed in Fig.~\ref{fig: FR Props  MHD}, all grow throughout the simulation but exhibit some significant variations. Overall, the more erratic evolution is consistent with the MHD model being generally more dynamic than the MFM. The most notable difference to the MFM relaxation MFR profiles is found in the average twist evolution. Despite the lack of driving, the MHD MFRs accumulated twist throughout the simulation, while for the MFM cases, the twist typically decreased.  As this accumulation starts already in the  early phases, it is therefore not tied to interactions with the boundaries of the modelling domain. Further inspection of the twist distribution within AR12473 MHD MFRs reveals that these MFRs are considerably less uniformly twisted than the MFRs in the MFM runs. This difference is particularly apparent when comparing with the AR11176 TMFM MFR, which also starts out with a high-twist region, enveloped by less twisted surroundings: While in the TMFM MFR, $T_w$ spreads to surrounding regions, the highly twisted regions in the MHD MFRs do not dissipate, but keep expanding on their own and maintaining their twist. This may ultimately explain the overall higher $T_w$ in the MHD cases compared to (T)MFM ones. 

For some MHD runs we found that parts of the initially coherent MFR structure became separated and started to evolve in a different manner. These regions were excluded in the extraction procedure, which ultimately leads to the jumps that are particularly pronounced in the MHD cross-section evolution around frame 25. Such cases could be explained by magnetic reconnection detaching part of the original structure. 

Despite the MFRs becoming partly detached and their cross-section growth stabilizing, they keep rising at least linearly throughout the simulation domain, even for the earlier relaxation times. Based on the behaviour and values of the investigated parameters, we can conclude that all MHD MFRs exhibit eruptive behaviour, despite notable transitions in the evolution curves between some runs (cf. shaded backgrounds in Fig.~\ref{fig: FR Props  MHD}). This is in contrast to the findings in \cite{Daei2023}, where earlier MHD relaxation times yielded non-eruptive MFRs. Like the MFM case, the latest three relaxation times produce MFRs with notably different properties and profiles than the earlier ones. They are larger and more twisted, rise faster and to higher altitudes, and reach higher values of decay index. The 'phase transitions' in the MHD MFR evolution curves are, however, more of quantitative nature, rather than qualitative like in the case of the MFM relaxation runs. In contrast to the MFM relaxation runs, the MHD runs also suggest that kink instability could have contributed in destabilising of some of the MHD MFRs (the reference value was exceeded both for the latest relaxation times and also some earlier times). However, the torus instability is still the most likely scenario which kept the MFR in an unstable configuration, given the attained values of $n$. 

As mentioned in Sect.~\ref{Sect: results}, the MHD MFRs' twist, decay index and cross-sectional areas do not necessarily reach higher values the later the relaxation time is chosen. Instead, the EOS properties show a local maximum at earlier relaxation times in Fig.~\ref{fig: FR Final Props}. The MHD EOS curves interestingly follow the same evolution as the average twist of the fully driven AR12473 TMFM MFR in the corresponding time window. Since the magnetic configurations of the TMFM MFR serve as initial conditions for the relaxation runs, it hints that the twist, which is present already in the MFR at relaxation, might be the governing parameter for MHD MFR evolution. We note, though, that the similarity between the TMFM twist curve and the MHD relaxation properties at EOS is approximate, and one would need further studies to confirm or disregard this theory. Additionally, for the erratic $T_w$ evolution, it does not seem clear if the ordering would change if the simulation time was expanded even further. 

Our analysis shows that MHD and MFM handle the same initial conditions very differently, and eruptivity in MHD does not necessarily translate into eruptivity in MFM. While differences are expected due to the different physics prescribed, the interplay of various effects is not a priori obvious. For example, MHD is generally evolving more dynamically, but on the other hand, it contains mass which needs to be moved in contrast to MFM. Moreover, for eruptive cases, the contributing instabilities may also change depending on the model. Furthermore, there are notable differences if driving is sustained beyond the point of eruptivity, as phase transitions in the MFR property profiles indicate. Our results thus highlight that care has to be taken both with the data-driving aspect and the actual modelling approach used for (data-driven) coronal magnetic field simulations. While it is clear that driving is critical for producing an eruption and that the duration of the driving affects MFR evolution, an important question remains: To what point is driving beyond eruptivity necessary to produce the most realistic MFRs, i.e. that matches best with the observations?  

\section{Conclusion}
\label{Sect: conclusion}

In the present work, we have investigated the effect of data driving on the evolution and eruptivity of magnetic flux ropes for two active regions, AR12473 and AR11176. We performed for both cases fully data-driven time-dependent magnetofrictional model \citep[][]{Pomoell19} runs and for the eruptive AR12473, we performed a series of relaxation runs using magnetofrictional as well as zero-beta magnetohydrodynamic approaches, where we systematically varied the relaxation times. 

The MFRs' evolution and eruptivity analysis was based on the evolution of geometrical MFR properties (cross-section and height) combined with stability properties (twist and decay index). Studying the evolution of characteristic MFR properties helps in assessing their eruptivity in the absence of exact knowledge of critical instability parameter thresholds (which are known only for idealised cases). In particular, exploring phase transitions and evolution of MFR key properties (increases of height, size, etc.) and instability parameters proved to be a powerful diagnostic tool for determining MFR eruptivity in our simulations. 

Our main findings are: 

\begin{enumerate}
    \item There are notable differences between the dynamics of the MFRs in the relaxation runs that use MFM and zero-beta MHD. Not only do they evolve differently (more erratic evolution and significant accumulation of twist in the MHD cases when compared to MFM), but the MHD MFRs were found to be considerably more eruptive despite starting from the same initial magnetic field configuration. 
    This is not unexpected due to the nature of the models. However, considering that both approaches are commonly used in magnetic flux rope studies, our results emphasize that in modelling context eruptivity is not necessarily a general property of some initial magnetic field configuration but can also be influenced by the physical prescriptions of the used model. 
    \item For the eruptive MFRs in the MFM relaxation runs we found that the twist metric $T_w$ was rather low and stagnating (between 0.9 and 1.1), while the decay index reached above 2 and beyond. This implies that the torus instability was the likely trigger for the eruptions. For the eruptive MHD cases, the torus instability also appears to be the likely trigger. However, some MFRs show significant twist in the magnetic field, making it difficult to rule out a contribution from the kink instability to the eruption. 
    \item Sustaining the driving beyond the point of eruptivity can significantly influence the evolution of MFRs. This can be seen by notable gaps and differences in the evolution profiles of MFR properties between sets of eruptive MFRs, relaxed at different times. This poses the question until which point driving must be continued to most accurately reproduce the physical reality of a solar eruption. 
\end{enumerate}

\begin{acknowledgements}
This work is part of the SWATNet project funded by the European Union’s Horizon 2020 research and innovation programme under the Marie Skłodowska-Curie grant agreement No 955620. DP acknowledges Horizon 2020 grant No 101004159 (SERPENTINE). RE thanks for the support received from NKFIH OTKA (Hungary, grant No. K142987). The NKFIH Excellence Grant TKP2021-NKTA-64 also supported this work.
RE is also grateful to STFC (UK, grant number ST/M000826/1) and PIFI (China, grant nr 2024PVA0043). SP acknowledges support from the projects C14/19/089  (C1 project Internal Funds KU Leuven), G0B5823N and G002523N (WEAVE)   (FWO-Vlaanderen), 4000145223 (SIDC Data Exploitation (SIDEX), ESA Prodex), and Belspo project B2/191/P1/SWiM. A.K.\ is supported by an appointment to the NASA Postdoctoral Program at the the NASA Goddard Space Flight Center (GSFC). JP and FD acknowledge funding from the Academy of Finland project SWATCH (343581).
\end{acknowledgements}

\bibliographystyle{aa}
\bibliography{main_clean}

\newpage

\appendix
\section{Peak twist and average $n$}
\label{Appendix}
\begin{figure}
     \centering     
     \includegraphics[width=\linewidth]{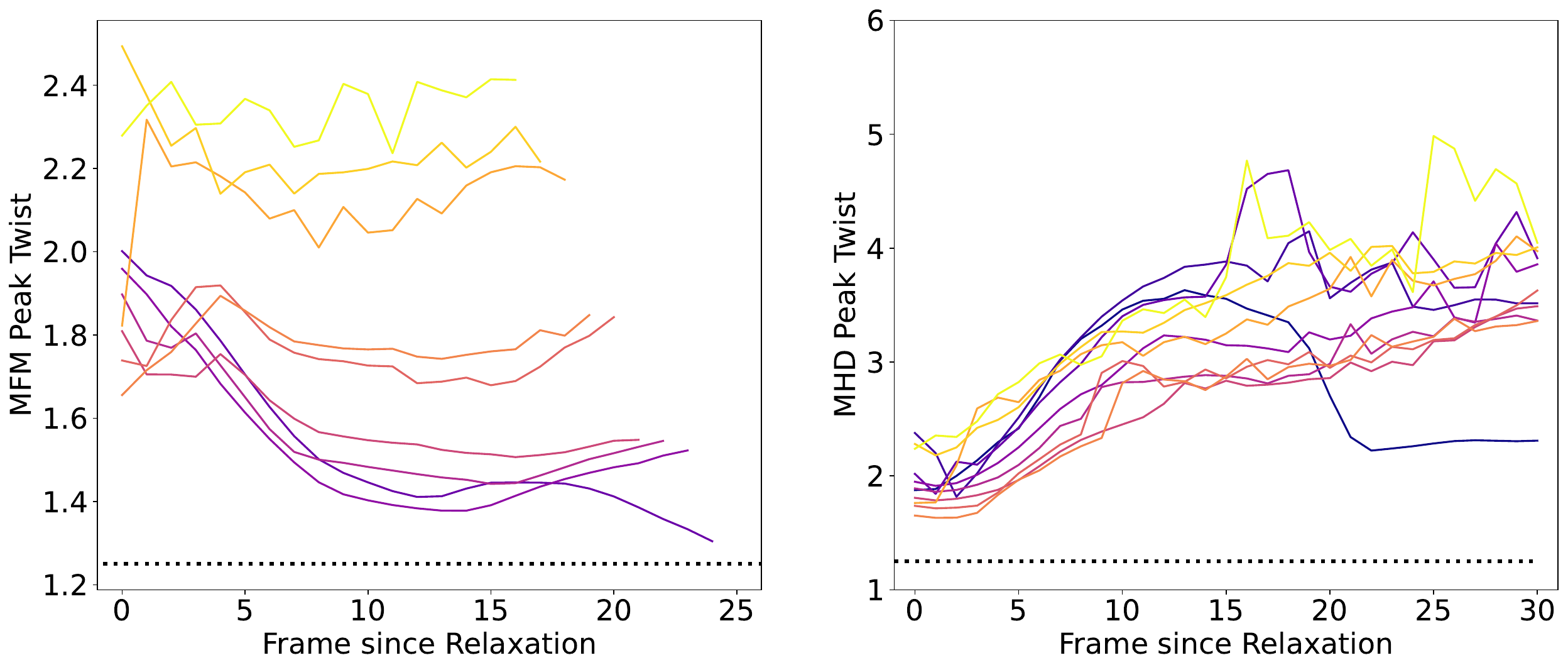}
     \caption{Peak $T_w$ for MFM and MHD relaxation simulation MFRs. Colouring as Fig~\ref{fig: FR Props MFM} and Fig~\ref{fig: FR Props  MHD}.}
     \label{fig: Peak Tw}
\end{figure}

To complement and solidify our analysis, we show the evolution of $T_w$ and $n$, when using different methods of obtaining said parameters. First, we calculate the peak twist within the relaxation MFRs in Fig.~\ref{fig: Peak Tw}. As expected, the values are higher and fluctuate notably stronger than in the averaged case. However, the separation of profiles are astonishingly similar in the MFM case, with the non-eruptive MFRs clustering at around $T_w^{max} \approx 1.5$, while the two sets of eruptive MFRs attain maximum twist values of about $T_w^{max} \approx 1.8$ and $T_w^{max} > 2$. The trends are also similar to the average $T_w$ trends, showing mostly declining and stagnating profiles. In the MHD relaxation MFRs, the peak twist values are, as in the average $T_w$ case, notably higher than for the MFM relaxation runs. They also exhibit no clear gaps, barring some outliers in the profiles of the earliest and latest relaxation runs. We note that the 29-12 MHD relaxation run had some outliers in the late stage of the simulation, which were removed in Fig.~\ref{fig: Peak Tw}. Similarly, as \cite{Duan2022}, we would argue that the more reliable indicator of kink instability is given when taking the twistedness of the whole MFR into account instead of a singular peak value, as it is not representative of the twist distribution of the full MFR structure. \\

\begin{figure}
     \centering     \includegraphics[width=\linewidth]{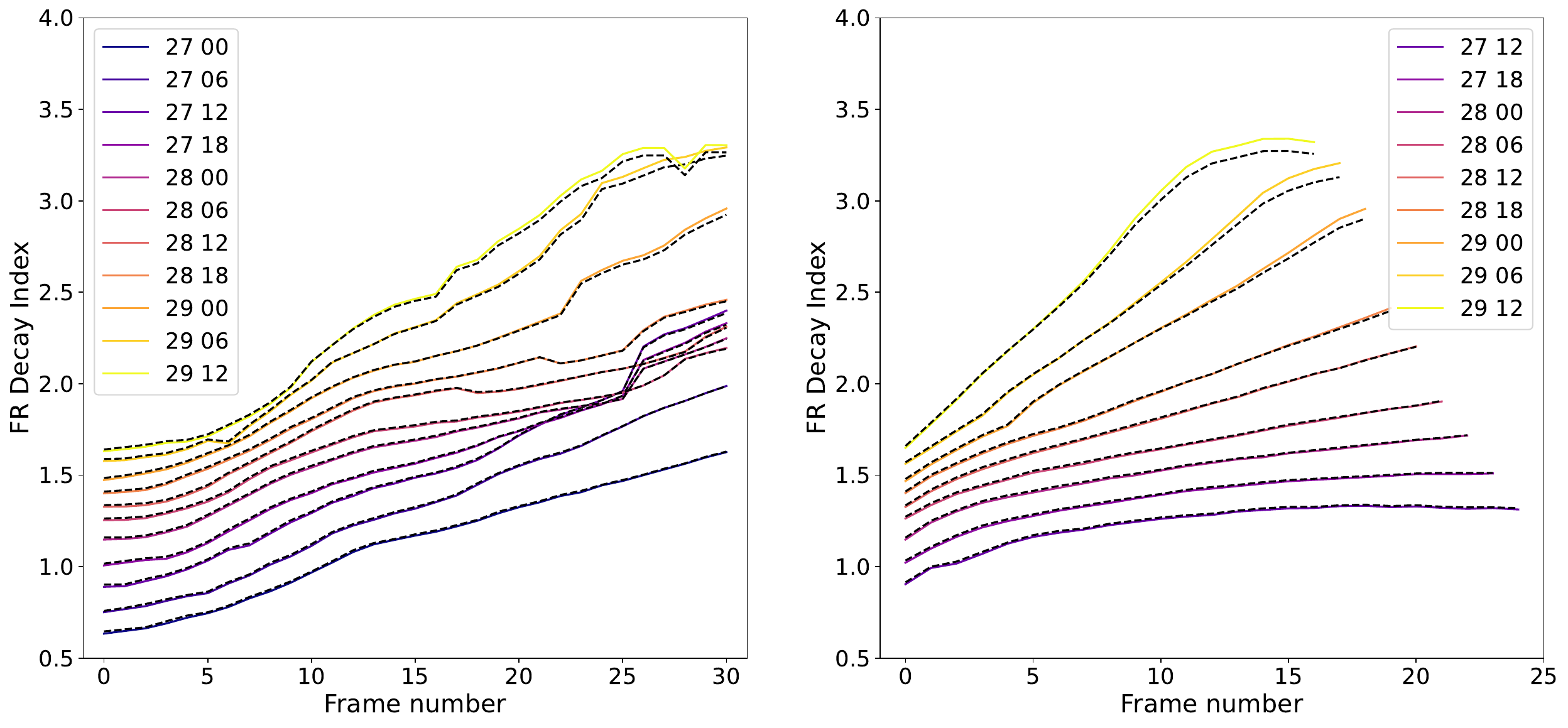}
     \caption{Comparison between decay index calculated at the location of the MFR centre (coloured lines) and calculated in a box around the MFR centre (dashed lines). Colouring as Fig~\ref{fig: FR Props MFM} and Fig~\ref{fig: FR Props  MHD}.}
     \label{fig: Decay Index Comparison}
\end{figure}

Finally, we show the differences between the decay index $n$ at a singular location, the MFR centre (coloured lines), and averaged in a box around this value (dashed lines) in Fig.~\ref{fig: Decay Index Comparison}. The box is chosen such that its length equals a fifth of the MFR cross-section length and height equals a fifth of the MFR cross-section height. The differences are negligible and only become visible in cases where the MFRs encounter the top of the modelling domain, namely the latest two relaxation times. For this reason, no further investigation on the averaged decay index has been taken as the differences are negligible. 

\end{document}